\documentclass{article}
\usepackage{icrctc07}
\DeclareSymbolFont{AMSb}{U}{msb}{m}{n}
\DeclareMathSymbol{\varnothing}{\mathord}{AMSb}{"3F}

\title{The reflecting surface of the MAGIC-II Telescope}
\shorttitle{MAGIC-II reflecting surface}
\authors{D.~Bastieri$^{1,2}$,
         J.~Arnold$^3$
         C.~Baixeras$^4$,
         O.~Citterio$^5$,
         F.~Dazzi$^{1,2}$,
         B.~De Lotto$^6$,
         M.~Doro$^{1,2}$,
         M.~Ghigo$^{7,5}$,
         E.~Giro$^{8,2}$,
         F.~Goebel$^9$,
         R.~Kosyra$^9$,
         E.~Lorenz$^{10}$,
         M.~Mariotti$^{1,2}$,
         R.~Mirzoyan$^9$,
         R.~Paoletti$^{11}$,
         G.~Pareschi$^{7,5}$,
         D.~Pascoli$^{1,2}$,
         A.~Pepato$^2$,
         L.~Peruzzo$^{1,2}$,
         A.~Saggion$^{1,2}$,
         P.~Sartori$^{1,2}$ and
         A.~Sillanp\"a\"a$^{12}$ for the MAGIC Collaboration$^{13}$}
\shortauthors{D.~Bastieri et al.}
\afiliations{$^1$Dipartimento di Fisica, Universit\`a di Padova, Via Marzolo 8, I-35131, Padova\\
             $^2$I.N.F.N.\ Sezione di Padova, Via Marzolo 8, I-35131, Padova\\
             $^3$LT Ultra GmbH, Wiesenstrasse 9, D-88634 Aftholderberg\\
             $^4$Departament de F\'\i sica, Universitat Aut\`onoma de Barcelona,
                 Edifici M, E-08193 Bellaterra\\
		 $^5$Media Lario Technologies S.r.l., Localit\`a Pascolo, I-23842, Bosisio Parini (LC)\\
             $^6$Dipartimento di Fisica, Univ.\ di Udine \& I.N.F.N. Trieste,
                 Viale delle Scienze 208, I-33100, Udine\\
             $^7$Osservatorio Astronomico di Brera, Via Brera 28, I-20121, Milano\\
             $^8$Osservatorio Astronomico di Padova, Vicolo Dell'Osservatorio 5, I-35122, Padova\\
             $^9$Max-Planck-Institut f\"ur Physik, F\"ohringer Ring 6, 80805, M\"unchen, Germany\\
             $^{10}$Institute for Particle Physics, Swiss Federal Institute of Technology (ETH), CH-8093 Z\"urich\\
             $^{11}$Dipartimento di Fisica, Universit\`a di Siena \& I.N.F.N. Sezione di Pisa,
                 Via Roma 56, I-53100, Siena
             $^{12}$Tuorla Observatory, V\"ais\"al\"antie 20, FI-21500 Piikki\"o\\
             $^{13}$Updated collaborator list at http://magic.mppmu.mpg.de/collaboration/members/index.html}
\email{denis.bastieri@pd.infn.it}

\abstract{The MAGIC Collaboration is building a second telescope, MAGIC II, improving the design of the current MAGIC Telescope.  MAGIC II is being built at $85\:\mathrm{m}$ of distance from MAGIC I, and will also feature a huge reflecting surface of $\sim240\:\mathrm{m}^2$ of area.  One of the improvement is the design for the mirror of MAGIC II, that are lighter and larger, being square of $1\:\mathrm{m}$ of side and weighting around $15\:\mathrm{kg}$.  For the development and production of the new mirrors, two different techniques, both reliable and affordable in price, were selected: the diamond milling of aluminium surfaces and the cold slumping of thin glass panes.  As tests for the second one are still ongoing, we present a description of the diamond milling technique, and its application and performance to the produced mirrors.}

\begin{document}
\maketitle

\section{Introduction}\label{sec:intro}
MAGIC the \emph{Major Atmospheric Gamma Imaging Cherenkov\/} Telescope, was designed to be \emph{the\/} Cherenkov telescope with the lowest envisaged energy threshold.  It was built by an international collaboration at the Roque de Los Muchachos ($2,200\,\mathrm{m}$ of altitude), a volcano rim in the island of La Palma, Canaries. Many technical developments were necessary to lower the energy threshold, but the expected scientific outcome was widely repaying the burdens of the construction.

The key to lower the energy threshold is increasing the area of the reflecting surface, as well as designing a trigger able to stand the high rate of the so called \emph{Night Sky Background}, likely to increase itself rapidly as more light is collected by the reflecting surface.  This is the reason why the MAGIC Telescope features the bigger reflecting surface ($236\:\mathrm{m}^2$) among other similar experiments.

In order to benefit from all the advantages of stereoscopic vision, the MAGIC Collaboration agreed on building MAGIC II\cite{magic2}, a new telescope, a \emph{clone}, at $85\:\mathrm{m}$ of distance from the old one.  The reflecting surface of MAGIC II is similar to the one of the older MAGIC, but with the experience gained in assembling mirrors for MAGIC I (square, with a side of $50\:\mathrm{cm}$)\cite{m1m}, we preferred to assemble larger mirrors for MAGIC II (still square, but with a side of $1\:\mathrm{m}$).

The huge reflector will still have an overall parabolic shape, which allows detected photons to keep the correct timing information, and is segmented into 236 smaller elements ($1\,\mathrm{m}\times 1\,\mathrm{m}$), each machined to spherical shape with the curvature radius that better fits the required parabolic shape.  Each element is an aluminium honeycomb core \emph{sandwiched\/} between two outer Al-layers using laminating adhesives.  The sandwich, called \emph{raw blank}, is later worked and polished with milling machines.  Details can be found in sec.\ \ref{sec:mirrors}, while the optical properties of the mirrors can also be found in sec.\ \ref{sec:optical}.

\section{The Mirrors}\label{sec:mirrors}
MAGIC II mirrors are a composite structure made up by a layer of AlMgSi1.0 $F=30$, an aluminum honeycomb and an outer aluminum box all glued together in a high pressure tank making up the so-called \emph{raw blank}.  Raw blanks are pre-shaped to spherical shape and then polished with a milling tool equipped with a diamond tip of \emph{large\/} $(\sim 1\:\mathrm{m})$ curvature radius.  The final curvature radius is the one that better matches the parabolic shape of MAGIC-II dish $(34.125\div 36.625\:\mathrm{m})$.  After diamond milling, front plates are coated with a hard, transparent protective layer against scratches and aging and the produced mirrors weight around $15\:\mathrm{kg}$.  Each mirror will be probably equipped with a heating system to prevent ice and dew formation.

Pre-shaping was first attempted for MAGIC-I mirrors in view of the construction of MAGIC-II mirrors.  MAGIC-I mirrors, in fact, are made up with a thicker slab of flat aluminium that is later premilled with an accuracy of better than $\frac{1}{10}\:\mathrm{mm}$.  Using pre-shaped raw-blanks, two major issues could be improved even for the smaller MAGIC-I mirrors:

\begin{itemize}
\item the thickness of the Al slab, needed for the milling, was reduced from 5 to $1\div 2\:\mathrm{mm}$;
\item \emph{premilling\/} could be skipped.
\end{itemize}

Pre-shaped mirrors are assembled, as the old one, in an \emph{autoclave\/} environment, but are lay onto a curved mould, that shapes the final raw blank with the requested curvature radius, between 34 and $36\:\mathrm{m}$.
\begin{figure*}[tbh]
\includegraphics[width=.45\textwidth]{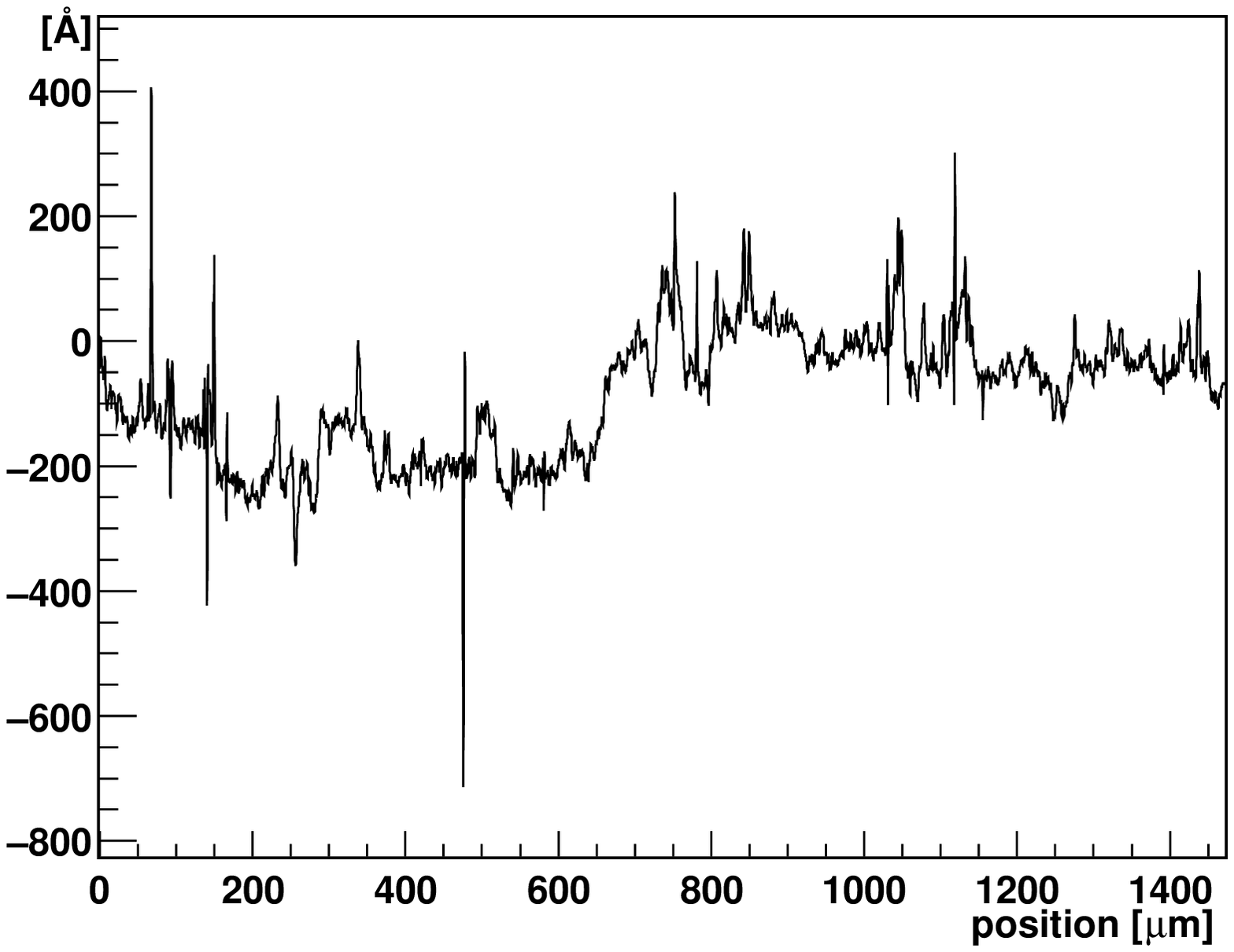}\hfill\includegraphics[width=.45\textwidth]{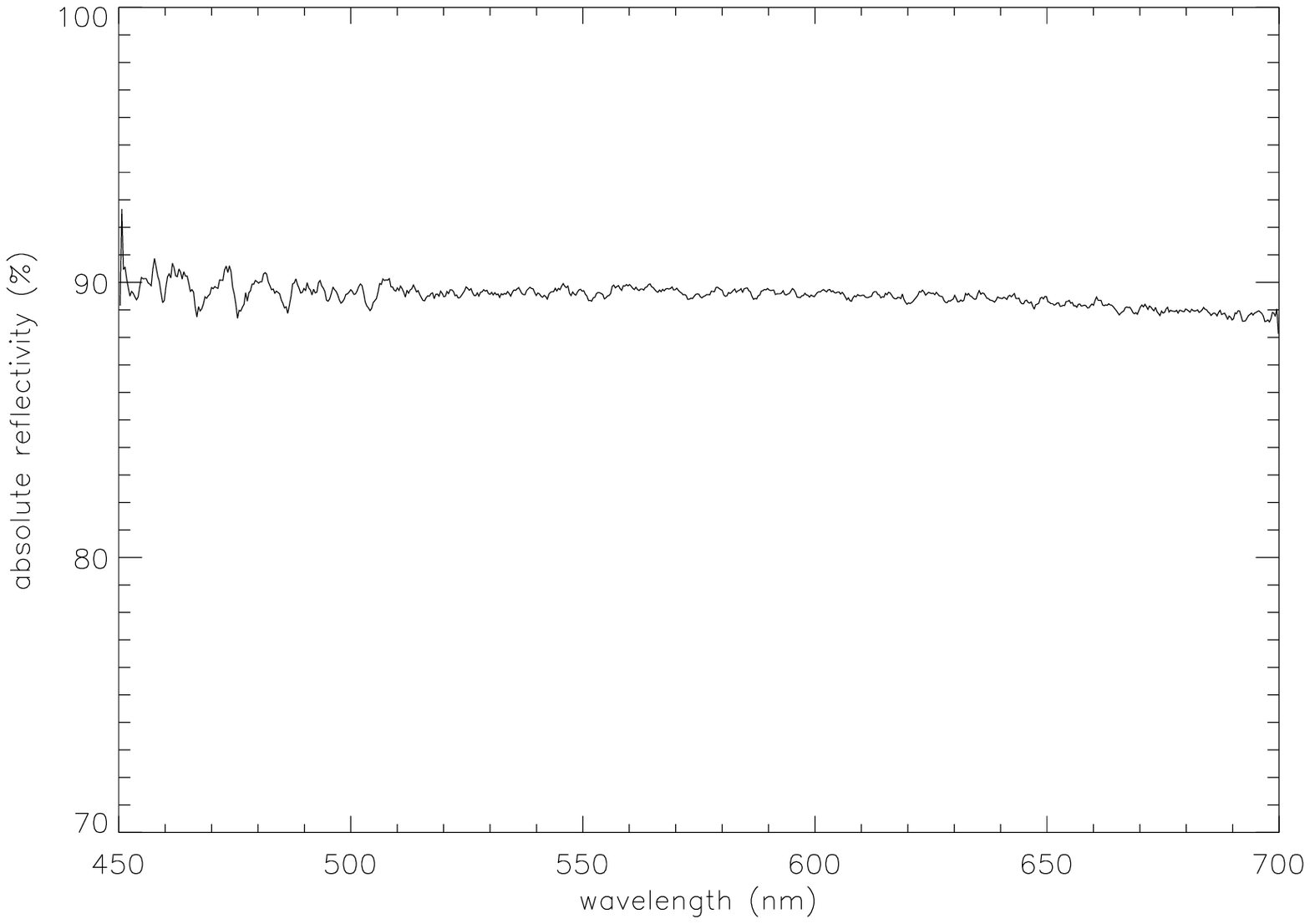}
\caption{\label{fig:roughness}\emph{(Left)\/} Roughness of a sample taken from a MAGIC mirror after milling and coating.  In the roughness diagram, the actual profile (measured in \AA ngstrom) is plotted against the position (in millimeters). \emph{(Right)\/} The typical reflectivity of a small portion of mirror before coating.}
\end{figure*}

Let us remind that, in this range, the \emph{sagittae\/} vary $\sim 0.5\:\mathrm{mm}$.  Therefore, we can produce all pre-shaped raw-blanks with just one \emph{gross\/} curvature radius and let the diamond milling machine refine them by removing just a minimal amount of material.  This results in a faster, and less expensive, overall procedure.

Working with thinner, but pre-shaped, Al-plates also makes the assembling and machining of larger mirrors easier: in fact, a $1\,\mathrm{m}\times 1\,\mathrm{m}$ spherical mirror of $\sim 34\:\mathrm{m}$ of curvature radius requests that $\sim 4\:\mathrm{mm}$ of material would have to be removed from its centre if it were assembled with a flat plate, whereas virtually no material at all is removed from pre-shaped mirrors.  Moreover, as MAGIC currently uses panels hosting four fixed mirrors each for active optics, increasing the mirror size also eliminates the necessity to use back-panels, as the mirrors themselves could be controlled with minor refinements to the actual active optics device.

Larger mirrors have nevertheless some drawbacks.  In fact, MAGIC is made up with many small spherical mirrors that best fit the desired overall parabolic shape: increasing mirror size makes the fit harder, at least for the outer mirrors, where the requested paraboloid differs more from a sphere. Astigmatic mirrors can adapt better to parabolic shapes, but their production can be quite difficult, and for MAGIC-II, if machining of astigmatic mirrors does not prove to be feasible via the diamond milling technique, it could be envisaged the construction of a mixed-size surface, with $1$-$\mathrm{m}$ mirrors in the inner rings and $50$-$\mathrm{cm}$ ones outside.

Coming back to raw blanks, they are composed of a 2-$\mathrm{mm}$ thick Al 3003 box, containing the Al 5052 honeycomb of $6.0\:\mathrm{cm}$ of thickness and sealed with the AlMgSi1.0 layer.  Three small aluminum plates, $12$-$\mathrm{mm}$ thick, are embedded into the honeycomb and glued to the outer box.  They host four screws each, to fix the finished mirror to the Active Mirror Control system of the telescope.  Final assembly of the raw blank parts is done using two layers of \textsf{3M} glue foils between box, honeycomb and front plate.  The gluing procedure consists of a curing process at $120^\circ$ and $5\:\mathrm{atm}$ of pressure.

The diamond milling of the surface is done by the \textsf{LT Ultra} company (Aftholderberg, Germany).  After diamond milling, the roughness of the surface is well below $10\:\mathrm{nm}\:rms$, as can be seen in fig.\ \ref{fig:roughness} for a typical profile analysed with a commercial surface roughness tester.  From the same picture one can also see one \emph{step\/} of the milling machine, that can follow the desired profile at a level of the micrometer.  Related to the roughness, the local reflectivity, lying between $85\%$ and $90\%$ in the visible band.

\section{Optical quality checks of the mirrors}\label{sec:optical}
\begin{figure*}
\begin{center}
\includegraphics[width=.4\textwidth]{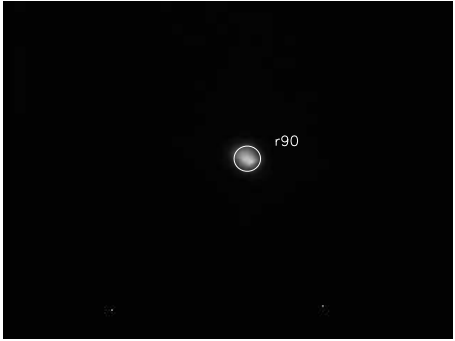}\hfill\includegraphics[width=.4\textwidth]{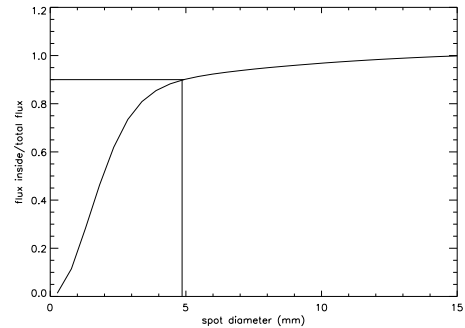}
\end{center}
\caption{\label{fig:dist}\emph{(Left)\/} \emph{Spot\/} appearance of a produced mirrors on the focal plane. \emph{(Right)\/} The distribution of the enclosed energy inside the \emph{spot}.}
\end{figure*}
To check the optical quality, we use an ultrabright blue LED that is reflected by the mirror under study onto a white screen: the reflected image, the \emph{spot}, is analysed with a CCD camera.  The centre of the screen and the LED are at a distance of $\sim 10\:\mathrm{cm}$, and are symmetric with respect to the mirror axis.  The distance between the mirror and the LED (and between the mirror and the screen) is equal to the nominal curvature radius (or twice the focal length) of the mirror itself, in such a way that a point image is reflected again into a point image.

The first batch of 12 mirrors underwent this kind of measurements, with results even beyond the expectations, as can be seen for the values of $R_{90}$, that is the radius of the circle, taken from the centre of gravity of the spot, containing 90\% of the total, reflected light.  As the picture is taken at twice the focal, when focusing light-rays coming from \emph{infinity\/} the spot is actually half the size of the measured one.  Looking at fig.\ \ref{fig:dist}, the result is that $90\%$ of the light from a parallel beam will be focused, on average, within a circle of $0.5\:\mathrm{cm}$ of diameter, or less than $\frac{1}{4}$ of the MAGIC pixel size (PMTs of $1"\varnothing$).

The effective radius of curvature is defined operatively as the distance between the spot and the mirror where the $R_{90}$ is minimum.  It is the effective radius of curvature that is taken into account for the correct positioning of the mirror onto the parabolic dish, having to match the local mean curvature radius of the paraboloid.

\section{Conclusions}
The reflecting surface of the MAGIC telescope is operating in open air already for some years.  During this period, few percent of it were damaged by water infiltrations inside the sandwich structure.  Due to the extremely bad weather conditions of winter 2004--2005, the continuous changing of state between water and ice had the effect of detaching some of the aluminium plates from the raw blanks structure.  An improved sealing was then devised in order to prevent water creeping inside the mirror structure.

On the other hand, even in these extreme weather conditions, with also strong wind blowing and \emph{calima\/} (Sahara sand particles with $5\div 10\:\mu\mathrm{m}\:\varnothing$), the hard surface seems to resist quite well: samples coming from mirrors that had to be substituted evidenced no change in local reflectivity.

Few mirrors made from pre-shaped raw-blanks are already installed on MAGIC and survived safely the same exceptionally hard winter.  Among them also some larger $1$-$\mathrm{m}$ mirror, mounted as a test for the new technology foreseen for MAGIC-II.

The technology of diamond milling applied to pre-shaped honeycombs proved reliable, and the results of the first batch even beyond expectations. While tests for the \emph{cold slumping\/} technique are advancing, the production of the first half of MAGIC-II mirrors with the diamond milling technique is going on, to met the schedule and have the new telescope ready for September 2008.

\section{Acknowledgements}
The production of the reflecting surface of the MAGIC Telescope was made possible by the financial contributions of the Italian INFN and INAF, the German BMBF, the Universitat Aut\`onoma de Barcelona and the Tuorla Observatory to whom goes our grateful acknowledgement.  Acknowledgements is also due to the LT Ultra company of Alt Holderberg for all their dedication in developing new technologies for the mirror production.  Finally, we would also like to thank the IAC for the excellent working conditions provided at El Roque de los Muchachos.

\end{document}